\documentclass[useAMS,referee]{biom_submission}
\usepackage{graphicx} 
\usepackage{setspace}
\usepackage{subfig}
\usepackage{amsmath}
\usepackage{amsfonts}
\usepackage{amssymb}
\usepackage{pgfplots}
\usepackage{xcolor}
\usepackage{tikz}
\usetikzlibrary{positioning}
\usepackage{listings}
\usepackage{booktabs}
\usepackage{multirow}
\usepackage{verbatim}
\usepackage{float}
\usepackage{appendix}
\usepackage{algorithm}
\usepackage{rotating}
\usepackage{booktabs}
\usepackage{threeparttable}
\usepackage{adjustbox}
\usepackage{pdflscape}
\usepackage{caption}

\usepackage{makecell}
\usepackage{longtable}
\usepackage{graphicx}
\usepackage{hhline}
\let\bibhang\relax
\usepackage{natbib}
\usepackage{comment}

\title[Risk-stratified case-control sampling and semiparametric weight calibration]{\parbox{0.9\textwidth}{\centering 
Risk-matched case-control sampling and semiparametric weight calibration for efficient biomarker assessment with binary outcomes
}}

\author{Jingyi Zhang$^{1}$, 
Fangya Mao$^{2*}$\email{fangyama@buffalo.edu}, and Lingxiao Wang$^{1**}$\email{lingxiao.wang@virginia.edu} \\
$^{1}$Department of Statistics, University of Virginia, Charlottesville, VA, USA \\
$^{2}$Department of Biostatistics, University at Buffalo, Buffalo, NY, USA}
\begin{document}


\date{{\it Received XXX} XXX. {\it Revised XXX} XXX.  {\it
Accepted XXX} XXX.}



\pagerange{\pageref{firstpage}--\pageref{lastpage}} 
\volume{XXX}
\pubyear{XXX}
\artmonth{XXX}


\doi{xxxx}


\label{firstpage}

\begin{abstract}
Two-phase designs provide a cost-effective framework for biomarker studies and have been widely used to reduce the burden of collecting expensive biomarker measurements. Most methodological research has focused on efficient Phase II sampling and estimation of biomarker--outcome associations. A natural subsequent step, however, is to incorporate identified biomarkers into risk prediction models, where measures of predictive performance become important inferential targets. Even for binary outcomes, methodological developments for two-phase designs that jointly address association parameters and predictive performance remain limited. Building on classical case-control and matched case-control designs developed primarily for association inference with binary outcomes, we propose a risk-matched case-control (RMCC) design that uses a scalar disease-risk score constructed from Phase I variables to define matching strata and guide Phase II sampling. To address the limitations of existing likelihood-based approaches, which typically require correct specification of the nuisance distribution of biomarkers, we develop a unified design-based inference framework based on inverse probability weighting for both association parameters and predictive performance measures, including positive predictive value and the area under the receiver operating characteristic curve. We further introduce target-specific weight scaling and calibration procedures and derive variance estimators that account for sampling variability in both phases. Simulation studies demonstrate the efficiency and robustness of the proposed design and inference procedures, and an application to a Taiwanese oral cancer cohort illustrates their potential utility.
\end{abstract}


\maketitle

\section{Introduction}

Two-phase designs arise naturally in observational studies where outcomes and basic confounders are available for an entire cohort, but additional covariates of primary interest are too costly to measure for all participants. Such settings are increasingly common in contemporary epidemiological and biomedical research. Examples include large-scale screening or surveillance programs in which standard demographic and clinical variables are collected for all individuals, while novel laboratory, genomic, or other specialized measurements are obtained only for selected participants; imaging studies where routine clinical data are available cohort-wide, but expert annotation or quantitative feature extraction is feasible only for a subset; and cohort studies in which registry, administrative, or electronic health record data are broadly available, whereas stored biospecimens can be assayed for molecular markers only in a subsample. By combining Phase I information with targeted Phase II measurements in a carefully selected subsample, two-phase designs provide a principled and cost-effective framework for evaluating biomarkers, including their associations with outcomes and their utility for risk prediction. 

For binary outcomes, the case-control (CC) design is one of the most commonly used two-phase designs in epidemiology and clinical research to assess risk factors for the disease of interest \citep{breslow1996,wacholder1992}. Under such designs, individuals in the Phase I cohort are stratified as cases (disease present) or controls (disease absent) and Phase II samples are drawn separately from these two strata, typically with different sampling fractions. For instance, in rare disease settings, all cases may be selected while only a random sample of controls is included. Matched case-control (MCC) designs further guide control selection by choosing controls with characteristics similar to those of the cases. Appropriately chosen matching factors can improve precision in estimating the marker–outcome associations \citep{kupper1981,sandoval2022}.

In practice, however, finding suitable controls becomes increasingly difficult as the number of matching variables grows, particularly when the variables are continuous or certain covariate combinations are uncommon. The choice of matching factors and their categories, as well as decisions to relax the matching criteria, for example, by broadening categories or matching on fewer variables when exact matches are unavailable, is often ad hoc and guided by convention rather than an explicit criterion linked to  inference about the target parameter.

Moreover, most existing work on MCC designs has focused on regression parameters characterizing marker–outcome associations. In many applications, scientific interest extends beyond characterizing marker--outcome associations to evaluating how well identified biomarkers predict disease risk, with measures such as positive predictive value (PPV) and the area under the receiver operating characteristic curve (AUC) serving as important targets of inference\citep{huang2016evaluating, pal2020, cao2023, mao2026}. 
It is challenging to appropriately evaluate the prediction performance of the Phase II risk model under conventional MCC designs because matching can alter distributions of the outcomes and predicted risks. Existing two-phase analysis for prediction performance  under likelihood-based frameworks can be applied to address selection bias but can be sensitive to model misspecification \citep{cao2023, mao2026}.

In contrast, under correct specification of the sampling mechanism, inverse probability weighted (IPW) and related design-based estimators converge to the same population limit as the  Phase I sample estimator, even when the outcome model is misspecified. However, they may be less efficient, particularly when Phase II inclusion probabilities vary substantially \citep{wang2020}. To improve efficiency, various weighting strategies, including calibration and raking, have been proposed \citep{deville1992,deville1993,breslow2009}. Nevertheless, IPW require that all Phase I individuals have positive probabilities of being selected into the Phase II sample, which can be violated under the standard MCC design when there exist unmatched controls. Moreover, the existing weighting adjustment strategies  focus on improving the marker-outcome association estimation, limiting their application to prediction performance evaluation. 

To address these challenges, we propose a unified framework for risk-matched case-control (RMCC) sampling and design-based inference for binary outcomes. The RMCC design uses a univariate disease-risk score estimated from Phase I data to define matching strata, guiding control selection for rare diseases and the selection of both cases and controls for more common diseases. The framework accommodates inferences for a range of target parameters , including both association parameters and predictive performance measures, such as PPV and AUC, for the expanded model incorporating expensive covariates. We develop IPW estimators and introduce a new class of weight adjustment strategies based on scaling and calibration adjustments tailored to specific target parameters. To support valid inference, we derive variance estimators that account for variability from both Phase I sampling from a large population and Phase II subsampling.

Our contributions are fourfold. First, we propose the use of scalar risk score to simplify matching between cases and controls. Second, we introduce a class of matched case-control designs that stratify the Phase I participants jointly by risk score and outcome, allowing the Phase II sampling design to accommodate multiple inferential targets while ensuring positive inclusion probabilities and oversampling individuals from more informative strata. 
Third, we develop robust and efficient design-based estimators for multiple inferential targets using target-specific weight adjustment strategies. Fourth, we provide inference procedures for regression parameters and predictive performance measures under general two-phase stratified designs, extending beyond the proposed RMCC design. 

The remainder of the paper is organized as follows. Section 2 defines the target parameters, and Section 3 introduces the RMCC sampling design. Section 4 develops the design-based inference framework and strategies for improving efficiency. Section 5 evaluates the proposed design and estimators through extensive simulations. Section 6 illustrates their application to data from a Taiwanese cohort study of oral cancer. Section 7 concludes with a discussion. Additional technical details are provided in the Supplementary Materials.

\section{Model and Targets of Estimation}\label{sec2}

We consider a two-phase sampling design in which the Phase I sample, denoted by $\mathcal{S}^{\mathrm{I}}$, consists of 
$N$ subjects randomly sampled from the study population. Let $Y$ denote a binary 
outcome, with $Y=1$ indicating a case and $Y=0$ indicating a control. Let 
$\pmb{X}_1$ denote the Phase I covariates, which are observed for all $N$ 
subjects, and let $\pmb{X}_2$ denote the Phase II covariates, which are measured 
only for a subsample selected on the basis of Phase I information. 
Conditional on $\pmb{X}_1$ and $\pmb{X}_2$, we posit the following logistic 
regression model for the binary outcome $Y$:
\begin{equation}\label{2.1}
    \log \frac{p(\pmb{x};\pmb{\beta})}{1-p(\pmb{x};\pmb{\beta})}
    = \pmb{\beta}^{\top}\pmb{x}
    = \beta_0 + \pmb{\beta}_1^{\top}\pmb{x}_1 + \pmb{\beta}_2^{\top}\pmb{x}_2,
\end{equation}
where $p(\pmb{x};\pmb{\beta}) = P(Y=1\mid \pmb{X}=\pmb{x})$,
$\pmb{X} = (1, \pmb{X}_1^{\top}, \pmb{X}_2^{\top})^{\top}$, and $\pmb{x}=(1,\pmb{x}_1^\top, \pmb{x}_2^\top)^\top$ is the realization of $\pmb{X}$. The parameter 
vector $\pmb{\beta} = (\beta_0, \pmb{\beta}_1^{\top}, \pmb{\beta}_2^{\top})^{\top}$ 
collects the regression coefficients, with $\pmb{\beta}_2$ representing the 
log-odds-ratio parameters associated with the Phase II covariates. Throughout, 
the superscript ``$\top$'' denotes the vector transpose.

We use $\mathcal{S}^{\mathrm{II}}$ to denote the Phase II sample which is a random sample of $\mathcal{S}^{\mathrm{I}}$. Let $\delta_i$ be the Phase II inclusion indicator for subject $i\in \mathcal{S}^{\mathrm{I}}$, where
$\delta_i = 1$ if $i\in \mathcal{S}^{\mathrm{II}}$ and has $\pmb{X}_{2i}$ observed; and $\delta_i = 0$ otherwise.
The observed two-phase data are then represented as
$\mathcal{D} = \{(Y_i, \pmb{X}_{1i}, \delta_i, \delta_i\pmb{X}_{2i}),\, i\in \mathcal{S}^{\mathrm{I}}\}$.
Since the Phase II selection mechanism $\{\delta_i,i\in \mathcal{S}^{\mathrm{I}}\}$ depends only on the Phase I data
$\mathcal{D}^{\mathrm{I}} = \{(Y_i, \pmb{X}_{1i}),\, i \in \mathcal{S}^{\mathrm{I}}\}$, the inclusion probability for $i\in\mathcal{S}^{\mathrm{I}}$ can be written as
\begin{equation}\label{2.2}
    \pi_i(\pmb{\rho}) = P(\delta_i = 1 \mid Y_i, \pmb{X}_{1i};\, \pmb{\rho}),
\end{equation}
where $\pmb{\rho}$ is a finite-dimensional parameter determined by the Phase II sampling design. For notation simplicity we drop $\pmb{\rho}$ from $\pi_i$.
Subject to a budget constraints on the overall Phase II sample size,
we impose the constraint
\begin{equation}\label{2.3}
    E\!\left(N^{-1}\sum_{i\in\mathcal{S}^{\mathrm{I}}}\delta_i\middle|\mathcal{D}^{\mathrm{I}}\right) = \sum_{i\in\mathcal{S}^{\mathrm{I}}}\pi_i=\tau \in (0,1),
\end{equation}
where $\tau$ is the target Phase II sampling fraction, and the $E(\cdot|\mathcal{D}^{\mathrm{I}})$ denotes the expectation with respect to the Phase II sampling indicator $\delta_i$ conditional on the Phase I sample.
Our goal is to develop efficient strategies for determining the Phase II sample inclusion probabilities $\{\pi_i,i\in\mathcal{S}^{\mathrm{I}}\}$, equivalently for randomly selecting a Phase II subsample $\mathcal{S}^{\mathrm{II}}$
of size $n$ with $E(n|\mathcal{D}^{\mathrm{I}})= \lfloor \tau N \rfloor$, based on the Phase I data
$\mathcal{D}^{\mathrm{I}}$, where $\lfloor a \rfloor$ denotes the greatest integer less than or equal to \(a\) . 

The majority of existing literature focuses on Phase II subsampling designs with the 
primary goal of improving estimation efficiency for $\pmb{\beta}_2$ using the 
resulting two-phase sample. 
In this work, we broaden the inferential scope to 
encompass the full parameter vector $\pmb{\beta}$ and relevant predictive 
performance metrics, specifically the positive predictive value (PPV) and the 
area under the receiver operating characteristic curve (AUC). 
Under model~\eqref{2.1} and for a prespecified risk threshold $c \in (0,1)$, the 
PPV is defined as the average predicted risk among subjects whose predicted risk 
exceeds $c$:
\begin{equation}\label{2.4}
    \mathrm{PPV}_c 
    = \frac{\displaystyle\sum_{i\in\mathcal{S}^{\mathrm{I}}} p(\pmb{X}_i;\pmb{\beta})\, 
            I\{p(\pmb{X}_i;\pmb{\beta}) > c\}}
           {\displaystyle\sum_{i\in\mathcal{S}^{\mathrm{I}}} I\{p(\pmb{X}_i;\pmb{\beta}) > c\}}.
\end{equation}
The AUC is defined as the probability that a randomly selected case has a higher 
predicted risk than a randomly selected control, with a value of $0.5$ assigned to 
ties \citep{gerds2008}. Specifically,
\begin{align}\label{2.5}
    \mathrm{AUC} 
    &= \frac{1}{N_1 N_0}\sum_{i\in\mathcal{S}^{\mathrm{I}}_1}\sum_{j\in\mathcal{S}^{\mathrm{I}}_0}\psi_{ij},
    \qquad
    \psi_{ij} = I\{p(\pmb{X}_i;\pmb{\beta}) > p(\pmb{X}_j)\}+\frac{1}{2}I\{p(\pmb{X}_i;\pmb{\beta}) = p(\pmb{X}_j\}
\end{align}
where $\mathcal{S}^\mathrm{I}_k=\{i:Y_i=k, i\in\mathcal{S}^{\mathrm{I}}\}$, $k=0, 1$, respectively represents the set of Phase I cases and controls, and  $N_1 = \sum_{i\in\mathcal{S}^{\mathrm{I}}} Y_i$ is the number of cases and $N_0 = N - N_1$ is 
the number of controls.

\section{A Risk-matched Case-Control Sampling Design}

We propose a flexible class of risk-matched case-control (RMCC) designs aimed at improving the efficiency of estimating multiple target parameters, including association parameters and risk parameters such as the positive predictive value (PPV) and area under the receiver operating characteristic curve (AUC), defined in Formulae (\ref{2.4}) and (\ref{2.5}), subject to the budget constraint in Formula (\ref{2.3}). The RMCC design is implemented in three steps.

\subsection*{Step 1: Develop a preliminary risk model to form matching strata in the Phase I sample}

Consider a working logistic regression model for $Y$ using only  the Phase I covariates $\pmb{X}_1$:
\begin{equation}\label{3.1}
    \operatorname{logit}\{P(Y=1\mid \pmb{X}_1;\pmb{\alpha})\}
    =\alpha_0 + \pmb{\alpha}_1^{\top}\pmb{X}_1,
\end{equation}
where $\pmb{\alpha} = (\alpha_0, \pmb{\alpha}_1^{\top})^{\top}$. 
Let $\widehat{\pmb{\alpha}}$ solve the Phase I sample estimating equation for the working mean model, defined by
\[
\sum_{i\in\mathcal{S}^{\mathrm{I}}}
\left\{
Y_i-p_i^{\mathrm{I}}(\pmb{\alpha})\right\}(1,\pmb{X}^\top_{1i})^\top=\pmb{0}
,
\]
where
$p_i^{\mathrm{I}}(\pmb{\alpha})
=
\operatorname{expit}\left(
\alpha_0+\pmb{\alpha}_1^{\top}\pmb{X}_{1i}
\right).$ 
The estimated risks are $\widehat{p}_i^{\mathrm{I}}=p_i^{\mathrm{I}}(\widehat{\pmb{\alpha}})$, for $i \in \mathcal{S}^{\mathrm{I}}$. 

We partition the Phase I sample into $H$ risk-based matching strata using according to these estimated risk $\widehat{p}_i^{\mathrm{I}}$, using user-specified cutpoints $0=c_0<c_1<c_2<\cdots < c_{H_1} <c_H=1$. 
Specifically, we define the matching stratum index as 
$$
\bar{p}_i^{\mathrm{I}}=h
\quad\text{if}\quad
c_{h-1}<\widehat{p}_i^{\mathrm{I}}\leq c_h,
\qquad h=1,\ldots,H.
$$ 
We further divide each matching stratum by case-control status $Y_i$, yielding $2H$ mutually exclusive sampling strata in the Phase I sample $\mathcal{S}^{\mathrm{I}}$, indexed by $(k,h)\in{0,1}\times{1,\ldots,H}$. The resulting strata are denoted by$\mathcal{S}^{\mathrm{I}}_{k,h}=\{i \in \mathcal{S}^{\mathrm{I}}:Y_i=k,\ \bar{p}_i^{\mathrm{I}}=h \}$ with corresponding stratum sizes $N_{k,h}=|\mathcal{S}_{k,h}^{\mathrm{I}}|$, for $k\in \{0,1\}$ and $h=1,\ldots,H$. 

The specified cutpoints should be chosen to ensure adequate numbers of subjects in each sampling stratum; in our implementation, we require $N_{k,h}\geq 5$. We recommend selecting the cutpoints based on percentiles of the fitted-risk distribution among cases rather than the full Phase I sample. This choice maintains broad overlap in the fitted-risk distributions of cases across the two phases, thereby preserving representation of the case risk spectrum and reducing reliance on extrapolation when estimating risk-based quantities. In our simulation and application studies, we use $H=3$ with cutpoints $c_1$ and $c_2$ specified as the tertiles of $\widehat{p}_i^{\mathrm{I}}$ among Phase I cases.

Given this stratification of the Phase I sample, the Phase II selection model in \eqref{2.3} can be written as
\[
    \pi_i(\pmb{\rho})
    = \sum_{h=1}^{3}\sum_{k=0}^{1}
      \mathbb{I}(i\in\mathcal{S}^{\mathrm{I}}_{k, h})\cdot\rho_{k,h},
\]
where  $\mathbb{I}\{\cdot\}$ indicates whether each Phase I individual $i$ is in the sampling stratum $\mathcal{S}^{\mathrm{I}}_{k,h}$, and $\pmb{\rho} = \{\rho_{k,h} :k \in \{0,1\}, h \in \{1,2,\ldots, H\}\},$ 
with $\rho_{k,h} = n_{k,h} / N_{k,h}$.

\subsection*{Step 2: Determine stratum-specific Phase II case allocation}

The RMCC design first determines the stratum-specific Phase II case allocation $n_{1,h}$, $(h=1, 2, \ldots, H)$ subject to $n_1=\sum_{h=1}^H n_{1,h}= \min\{ \lfloor n/2 \rfloor, N_1\}$ This rule selects all available cases when they constitute fewer than half of the planned Phase II sample $n$ and otherwise allocates half of the sample of $n$ to cases for one-to-one case-control matching. 
For each case-sampling stratum $\mathcal{S}^{\mathrm{I}}_{1,h}$, define the mean absolute residual
\[
    A_{1,h} = \frac{1}{N_{1,h}}\sum_{i \in \mathcal{S}^{\mathrm{I}}_{1,h}} |r_i^{\mathrm{I}}|,
    \qquad r_i^{\mathrm{I}} = Y_i - \widehat{p}_i^{\mathrm{I}}.
\]
The $n_1$ cases are initially allocated across strata in proportion to
$A_{1,h}$, subject to the capacity constraints $n_{1,h} \leq N_{1,h}$.
If the  allocation for any stratum exceeds its capacity
$N_{1,h}$, its allocation is set to be $N_{1,h}$. The remaining sample
size, obtained by subtracting the allocations fixed at their capacities
from $n_1$, is then redistributed among the uncapped strata in proportion
to their respective $A_{1,h}$ values. This capping and redistribution
procedure is repeated until all $n_1$ cases are allocated and no stratum
exceeds its capacity.

\subsection*{Step 3: Determine stratum-specific Phase II control allocation}

We then allocate $n_0=n-n_1=\sum_{h=1}^H n_{0, h}$ controls across the $H$ control strata $\mathcal{S}^{\mathrm{I}}_{0,h}, h=1, 2, \ldots, H$. 
The initial control allocation is proportional to the corresponding case allocations, $n_{0,h} \propto n_{1,h}$, subject to the capacity constraints $n_{0,h} \leq N_{0,h}$, for $h=1,2, \ldots, H$. 
If a proportional allocation exceeds $N_{0,h}$ for any stratum, that stratum is capped at $N_{0,h}$, and the remaining control units are redistributed among the uncapped strata. To preserve risk matching, the redistribution prioritizes uncapped strata whose risk categories are closest to those of the capped strata. The process continues until all $n_0$ controls are allocated. 
After determining the allocations $\{n_{k, h},k=0,1, h=1,2, \ldots, H\}$,  
we select $n_{k,h}$ subjects from each $\mathcal{S}_{k,h}^{\mathrm{I}}$ by simple random sampling without replacement (SRSWOR), yielding the Phase II samples $\mathcal{S}_{k,h}^{\mathrm{II}}$.

When cases are rare and all are included in Phase II, \(n_{1,h}=N_{1,h}\approx N_1/H\). 
Allocating controls in proportion to the case counts then yields \(n_{0, h}\approx(n-N_1)/H\), corresponding to approximately equal allocation across the three risk-matching strata within each outcome group. 
When \(\pmb{X}_1\) is predictive of \(Y\), however, we typically have \(N_{0,1}>N_{0,2}>\cdots>N_{0,H}\) and \(\rho_{0,1}<\rho_{0,2}<\cdots <\rho_{0,H}\) by using  percentiles of the fitted-risk distribution among case as the cutoffs for the risk strata. 
Thus, RMCC preferentially samples controls with higher fitted risks and, consequently, larger absolute residuals $|r_i^{\mathrm{I}}|$. 
When cases are more common and subsampled, the residual-based case allocation combined with risk-matched control allocation similarly favors subjects with larger absolute residuals.

The RMCC design can therefore be viewed as a smoothed variant of the local case-control sampling design of \citet{wang2020}, In their design, individual inclusion probabilities are proportional to $|r_i^{\mathrm{I}}|$ and can vary substantially, potentially reducing the efficiency of IPW estimators.
In RMCC, residuals instead guide allocation \emph{across} sampling strata rather than directly determining individual inclusion probabilities. Subjects within a stratum share a common inclusion probability, while strata with larger mean absolute residuals receive larger allocations. By retaining subjects across the full range of risk-matching strata, RMCC provides broad coverage of the joint distribution of $(Y,\pmb{X}_1)$ while avoiding extremely small individual inclusion probabilities. This feature can improve the stability of inverse-probability-weighted estimators of $\pmb{\beta}$, the PPV, and the AUC.

\section{Estimation and Inference}

\subsection{Inverse Probability Weighted Estimation and Weight Calibration for $\pmb{\beta}$}

For estimating $\pmb{\beta}$, the inverse probability weighting (IPW) approach accounts for the 
unequal Phase II inclusion probabilities by assigning each sampled 
subject the weight $w_i = \pi_i^{-1}(\pmb{\rho})$. 
Under the analysis model (\ref{2.1}), the IPW estimating 
equation for $\pmb{\beta}$ is
\[
    U(\pmb{\beta})
    = \sum_{i\in\mathcal{S}^{\mathrm{I}}} \delta_i w_i
      \{Y_i - p(\pmb{X}_i;\pmb{\beta})\}\pmb{X}_i = \pmb{0},
\]
where $\delta_i$ indicates whether $i\in \mathcal{S}^\mathrm{I}$ is selected into $\mathcal{S}^{\mathrm{II}}$ (i.e., $\delta_i=1$ if $i\in \mathcal{S}^{\mathrm{II}}$, and $\delta_i=0$ if $i\in \mathcal{S}^{\mathrm{I}}-\mathcal{S}^{\mathrm{II}}$), with the expectation $E(\delta_i)=\pi_i(\pmb{\rho})$. We use $\widehat{\pmb{\beta}}^{\scriptscriptstyle\mathrm{IPW}}$ to denote the solution of $U(\pmb{\beta})=\pmb{0}$ for $\pmb{\beta}$. 

Although IPW estimators are design-consistent under a correctly specified 
selection model $\pi(\pmb{\rho})$, they can be inefficient, particularly when the Phase II 
sample is small or selection probabilities vary substantially across 
subjects. Efficiency can be improved by incorporating the Phase I 
information $\mathcal{D}^{\mathrm{I}}$ through calibrated inverse 
probability weighting \citep{lumley2011, shin2020}. The basic idea is 
to modify the IPW weights so that weighted summaries computed from the 
Phase II subsample reproduce corresponding summaries from the full Phase I 
cohort, thereby reducing the variability induced by Phase II subsampling 
while preserving design consistency.

We propose a two-step procedure that combines model-assisted weight 
calibration with stratum-specific weight scaling. 
The first step calibrates the IPW weights using information available for all Phase I 
subjects; the second step stabilizes the calibrated weights within the 
risk-outcome strata of the RMCC design. 
Because stratum-level scaling alters the weighted case-control composition, a stratum-specific offset 
is incorporated into the estimating equation to account for this change.

\subsubsection{Calibration step}

We use $\pmb{u}_i^{\mathrm{I}} = \{Y_i - \widehat{p}_i^{\mathrm{I}}\}
\pmb{X}_{1i}$ as the calibration variable, where $\widehat{p}_i^{\mathrm{I}}=p(\pmb{X}_{1i},\widehat{\pmb{\alpha}})$ 
is the predicted risk obtained from the preliminary risk model~\eqref{3.1} with $\widehat{\pmb{\alpha}}$ being the estimated regression parameters of model~\eqref{3.1}. Let $F_i$ denote 
the adjustment factor applied to the original IPW weight $w_i$ for $i\in\mathcal{S}^{\mathrm{II}}$. We 
determine $\{F_i,i\in\mathcal{S}^{\mathrm{II}}\}$ by requiring the calibrated sum of Phase II scores equals the Phase I score total:
\begin{equation}\label{4.1}
    \sum_{i\in\mathcal{S}^{\mathrm{I}}} \delta_i F_i w_i \pmb{u}_i^{\mathrm{I}}
    = \sum_{i\in\mathcal{S}^{\mathrm{I}}} \pmb{u}_i^{\mathrm{I}}.
\end{equation}
Minimising the chi-square distance between the original and calibrated 
weights subject to \eqref{4.1} yields the linear calibration factor 
$F_i = 1 + \widehat{\pmb{\eta}}^{\top}\pmb{u}_i^{\mathrm{I}}$, where
\[
    \widehat{\pmb{\eta}}
    = \left(\sum_{i\in\mathcal{S}^{\mathrm{I}}} \delta_i w_i \pmb{u}_i^{\mathrm{I}}\pmb{u}_i^{\mathrm{I}\top}
      \right)^{-1}
      \left\{\sum_{i\in\mathcal{S}^{\mathrm{I}}}\pmb{u}_i^{\mathrm{I}}
           - \sum_{i\in\mathcal{S}^{\mathrm{I}}}\delta_i w_i\pmb{u}_i^{\mathrm{I}}\right\}.
\]
The calibrated IPW weight for $i\in\mathcal{S}^{\mathrm{I}}$ is denoted by ${w}_i^{(c)} = F_i w_i$.

This calibration step exploits the Phase I information on $(Y, \pmb{X}_1)$ 
and can improve estimation efficiency for $(\beta_0, \pmb{\beta}_1)$. Under suitable 
regularity conditions, it retains design consistency and has properties 
analogous to generalized raking estimators \citep{lumley2011, wang2025}. 
However, since $\pmb{u}_i^{\mathrm{I}}$ contains no information on $\pmb{X}_2$ conditional on $Y$ and $\pmb{X}_1$, 
this step alone cannot improve efficiency in estimating $\pmb{\beta}_2$, 
nor does it control the weight variation induced by differential 
sampling fractions across risk-outcome strata, particularly between the cases and controls.

\subsubsection{Stratum-specific scaling and semiparametric adjustment}

To further stabilise the calibrated weights, we apply stratum-specific 
scaling. For $i\in \mathcal{S}^{\mathrm{II}}_{k,h}$, define the scaled weight
\[
    \widetilde{w}_i = \rho_{k,h} \cdot F_i w_i
    = \frac{n_{k,h}}{N_{k,h}} \cdot F_i w_i,
    \qquad i \in\mathcal{S}^{\mathrm{II}}_{k,h},
\]
This scaling reduces 
between-stratum weight variation. However, because RMCC samples cases 
and controls at different rates across risk strata, the scaling alters 
the weighted case--control composition within each stratum; directly 
solving the standard weighted logistic estimating equation with 
$\widetilde{w}_i$ would consequently distort estimation of $\beta_0$ 
and potentially the coefficients associated with $\pmb{X}_1$.

We therefore introduce a stratum-specific offset to restore the 
sampling-induced case--control odds. For subject $i$ in risk-matching stratum 
$h$, define
\[
    \widetilde{p}_i(\pmb{\beta})
    = \operatorname{expit}\!\left[
        \pmb{\beta}^{\top}\pmb{X}_i
        + \log\!\left\{
            \frac{n_{1,h} N_{0,h}}{n_{0,h} N_{1,h}}
          \right\}
      \right],
    \qquad i \in \mathcal{S}^{\mathrm{II}}_{k,h}.
\]
The semiparametric calibrated IPW (SCIPW) estimator 
$\widehat{\pmb{\beta}}^{\scriptscriptstyle\mathrm{SCIPW}}$ is then obtained 
by solving
\begin{equation}\label{4.2}
    U^{\scriptscriptstyle\mathrm{SCIPW}}(\pmb{\beta})
    = \sum_{i \in \mathcal{S}^{\mathrm{II}}}
      \widetilde{w}_i\,
      \{Y_i - \widetilde{p}_i(\pmb{\beta})\}\,
      \pmb{X}_i
    = \pmb{0}.
\end{equation}
The offset restores the target case--control odds within each stratum 
after weight scaling, so that the SCIPW procedure jointly achieves 
model-assisted efficiency gains and weight stabilization without 
sacrificing consistency. 

\noindent {\bf Theorem 1.}
Under regularity conditions \textbf{C1--C8} listed in  Supplementary Materials Section 1, and assuming that $n/N \rightarrow \tau$ for some $0<\tau<1$ as $N,n\rightarrow\infty$, the SCIPW estimator is consistent and asymptotically normal when the full logistic regression model is correctly specified:
$$
\widehat{\pmb{\beta}}^{\scriptscriptstyle\mathrm{SCIPW}}
\xrightarrow{p}
\pmb{\beta},
\qquad
\sqrt{n}
\left\{
\widehat{\pmb{\beta}}^{\scriptscriptstyle\mathrm{SCIPW}}-\pmb{\beta}
\right\}
\xrightarrow{d}
N\left\{
\pmb{0},
\pmb{\Sigma}^{\scriptscriptstyle\mathrm{SCIPW}}
\right\}.
$$
Consequently, $\widehat{\pmb{\beta}}^{\scriptscriptstyle\mathrm{SCIPW}}
=\pmb{\beta}+O_p(n^{-1/2})$ (proof in Supplementary Materials Section 2).

\subsection{Estimator for the PPV}

Let $\widehat{p}_i^{\scriptscriptstyle\mathrm{SCIPW}} = 
\operatorname{expit}(\pmb{X}_i^{\top}
\widehat{\pmb{\beta}}^{\scriptscriptstyle\mathrm{SCIPW}})$ denote the 
fitted risk for subject $i \in \mathcal{S}^{\mathrm{II}}$. The SCIPW estimator of 
$\mathrm{PPV}_c$ defined in \eqref{2.4} is
\begin{equation}\label{4.3}
    \widehat{\mathrm{PPV}}_c^{\scriptscriptstyle\mathrm{SCIPW}}
    = \frac{\displaystyle\sum_{i \in \mathcal{S}^{\mathrm{II}}} w_i\,
            \widehat{p}_i^{\scriptscriptstyle\mathrm{SCIPW}}\,
            I\!\left(\widehat{p}_i^{\scriptscriptstyle\mathrm{SCIPW}} > c\right)}
           {\displaystyle\sum_{i \in \mathcal{S}^{\mathrm{II}}} w_i\,
            I\!\left(\widehat{p}_i^{\scriptscriptstyle\mathrm{SCIPW}} > c\right)}.
\end{equation}

\subsection{Estimator for the AUC}

The IPW estimator of the AUC defined in Formula \eqref{2.5} is
\[
    \widehat{\mathrm{AUC}}^{\scriptscriptstyle\mathrm{IPW}}
    = \frac{1}{N_1 N_0}
      \sum_{i \in \mathcal{S}^{\mathrm{II}}_{1}}\sum_{j \in \mathcal{S}^{\mathrm{II}}_{0}}
      w_{ij}\,\widehat{\psi}_{ij},
\]
where $\mathcal{S}^{\mathrm{II}}_{1}$ and $\mathcal{S}^{\mathrm{II}}_{0}$ are the sets of Phase II cases 
and controls, $w_{ij} = \pi_{ij}^{-1}$ is the 
inverse of the joint inclusion probability 
$\pi_{ij} = P(\delta_i \delta_j = 1)$, and $\widehat{\psi}_{ij}$ is the comparison score computed from IPW-estimated risks \citep{yao2015}. Under the zero-gradient condition and suitable regularity conditions, coefficient estimation has no first-order effect on AUC variability (Supplementary Materials Section 3). Thus, replacing IPW with SCIPW coefficient estimates does not change the first-order variance when both converge to the same limit.

We therefore directly adjust the pairwise weights so that the Phase II AUC matches the full Phase I AUC under the same preliminary risk model (\ref{3.1}). Let $f_{ij}$ be a 
pairwise calibration adjustment factor determined by requiring the 
weighted Phase II case--control comparison scores to reproduce the 
corresponding Phase I total under the preliminary model~\eqref{3.1}:
\begin{equation}\label{4.4}
    \sum_{i \in \mathcal{S}^{\mathrm{I}}_{1}}\sum_{j \in \mathcal{S}^{\mathrm{II}}_{0}}
    w_{ij} f_{ij}\,\psi_{ij}^{\mathrm{I}}
    = \sum_{i\in\mathcal{S}^{\mathrm{I}}_1}\sum_{j\in\mathcal{S}^{\mathrm{I}}_0} \psi_{ij}^{\mathrm{I}},
\end{equation}
where $\psi_{ij}^{\mathrm{I}}$ is the comparison score computed from the 
preliminary fitted risks $\widehat{p}_i^{\mathrm{I}}$. The calibrated IPW estimator 
of the AUC is then can be obtained as follows
\begin{equation}\label{4.5}
    \widehat{\mathrm{AUC}}^{\scriptscriptstyle\mathrm{CIPW}}
    = \frac{1}{N_1 N_0}
      \sum_{i \in \mathcal{S}^\mathrm{II}_{1}}\sum_{j \in \mathcal{S}^\mathrm{II}_{0}}
      f_{ij} w_{ij}\,\widehat{\psi}_{ij}.
\end{equation}
We note that with the linear pairwise calibration factor $f_{ij}=1+\hat{\eta}\psi_{ij}^I$, the calibration estimator $\widehat{AUC}^{CIPW}$ admits the exact augmentation representation
\begin{equation}\label{4.5}
    \widehat{\mathrm{AUC}}^{\scriptscriptstyle\mathrm{CIPW}}
    = \widehat{AUC}^{\scriptscriptstyle\mathrm{IPW}}+\widehat{\gamma}\left(AUC^I-\widehat{AUC}^{\scriptscriptstyle\mathrm{I, IPW}}\right),
\end{equation}
where $AUC^I$ and $\widehat{AUC}^{\scriptscriptstyle\mathrm{I, IPW}}$ \(\widehat{\mathrm{AUC}}^{\mathrm{I,IPW}}\) evaluate the same fitted preliminary risk model using the full Phase I cohort and the IPW Phase II sample, respectively. The augmentation coefficient $\widehat{\gamma}=\left\{\sum_{i\in\mathcal{S}^\mathrm{II}_{1}, j\in\mathcal{S}^\mathrm{II}_{0}}w_{ij}\left(\psi_{ij}^I\right)^2\right\}^{-1}\left(\sum_{i\in\mathcal{S}^\mathrm{II}_{1}, j\in\mathcal{S}^\mathrm{II}_{0}}w_{ij}\psi_{ij}^I\widehat{\psi}_{ij}\right)$ captures the weighted association between the preliminary- and full-model pairwise comparison scores, \(\psi_{ij}^{\mathrm I}\) and \(\widehat\psi_{ij}\), through a regression fitted without an intercept. 

 \noindent\textbf{Theorem 2.}
Under C1--C3 and C9 in the Supplementary Materials,
\[
\widehat{\mathrm{AUC}}^{\scriptscriptstyle\mathrm{CIPW}}
-\widehat{\mathrm{AUC}}^{\mathrm{full}}\xrightarrow{p}0.
\]
If C10 also holds, then
\[
\sqrt N\left\{
\widehat{\mathrm{AUC}}^{\scriptscriptstyle\mathrm{CIPW}}-\theta_\ast\right\}
\xrightarrow{d}N\left(0,\Sigma_{\mathrm{AUC}}^{\scriptscriptstyle\mathrm{CIPW}}\right),
\qquad
\Sigma_{\mathrm{AUC}}^{\scriptscriptstyle\mathrm{CIPW}}
=V_{\mathrm I}+V_{\mathrm{II}},
\]
where $V_{\mathrm I}$ and $V_{\mathrm{II}}$ are the Phase I and Phase II variance components defined in Supplementary Materials Section 3 ((proof in Supplementary Materials Section 3). 

\subsection{Influence Function Based Variance Estimation for $\widehat{\pmb{\beta}}^\text{SCIPW}$ and $\widehat{\mathrm{AUC}}^\text{CIPW}$}
The variance of $\widehat{\beta}^\text{SCIPW}$ can be decomposed using the total variance formula as follows:
\begin{align}
Var(\widehat{\pmb{\beta}}^\text{SCIPW}) 
=Var \left\{ E \left( \sum_{i \in \mathcal{S}^\mathrm{I}} \delta_i \widetilde{w}_i\Delta_i(\pmb{\beta})\middle|\mathcal{S}^\mathrm{I} \right) \right\}
+ E\left\{ Var \left( \sum_{i \in \mathcal{S}^\mathrm{I}} \delta_i\widetilde{w}_i \Delta_i(\pmb{\beta}) \middle|\mathcal{S}^\mathrm{I}\right) \right\},
\label{var_scipw}
\end{align}
where 
$E(\cdot|\mathcal{S}^\mathrm{I})$ and $Var(\cdot|\mathcal{S}^\mathrm{I})$ 
are the expectation and variance with respect to the random sampling of $\mathcal{S}^\mathrm{II}$ from $\mathcal{S}^\mathrm{I}$ (i.e., $\{\delta_i,i\in \mathcal{S}^\mathrm{I}\}$). 
$E(\cdot)$ and $Var(\cdot)$ are with respect to the randomness of $\mathcal{S}^\mathrm{I}$.
The first summand of (10) can be estimated by the approximately unbiased estimator $v_1=N(N-1)^{-1}\sum_{i \in \mathcal{S}^\mathrm{II}}w_i\Delta_i^{\otimes2}\left(\widehat{\pmb{\beta}}^\text{IPW}\right)$, where $\Delta_i(\widehat{\pmb{\beta}}^\text{IPW})=\partial \widehat{\pmb{\beta}}^\text{IPW}/\partial w_i$ is the sample influence function of $\widehat{\pmb{\beta}}^\text{IPW}$, and $\pmb{a}^{\otimes2}=\pmb{a}\pmb{a}^\top$ represents the outer product operator of the vector $\pmb{a}$.
The second summand of (\ref{var_scipw}) can be estimated by the approximately unbiased estimator $v_2=var\left\{ \sum_{i \in \mathcal{S}^\mathrm{I}} \delta_i w_i \Delta_i \left(\widehat{\pmb{\beta}}^\text{SCIPW}\right)\middle|\mathcal{S}^\mathrm{I}\right\}$ , where $\Delta_i(\widehat{\pmb{\beta}}^\text{SCIPW})=\sum_{j\in \mathcal{S}^\mathrm{II}}\partial\widehat{\pmb{\beta}}^\text{SCIPW}/\partial F_j\cdot \partial F_j/\partial w_i+\partial\widehat{\pmb{\beta}}^\text{SCIPW}/\partial w_i$, with the two terms accounting, respectively, for the randomness due to the generation of $\{F_i,i\in \mathcal{S}^\mathrm{II}\}$ and final estimation of $\pmb{\beta}$. Combining $v_1$ and $v_2$, we estimate the variance of $\widehat{\pmb{\beta}}^\text{SCIPW}$ by:
\begin{equation}
    var\left(\widehat{\pmb{\beta}}^\text{SCIPW}\right)=\frac{N}{N-1} \sum_{i \in \mathcal{S}^\mathrm{II}}w_i \Delta_i^{\otimes 2} \left(\widehat{\pmb{\beta}}^\text{IPW}\right)+\sum^1_{k=0}\sum_{h=1}^3(1-f_{k,h})n_{k,h}{\widehat{\pmb{\sigma}}}^{2}_{k,h},
\end{equation}
where $f_{k,h}=n_{k,h}/N_{k,h}$, and $\widehat{\pmb{\sigma}}^{2}_{k,h}=(n_{k,h}-1)^{-1}\sum_{i \in\mathcal{S}^\mathrm{II}_{k,h}} \left\{w_{i}\Delta_i\left(\widehat{\pmb{\beta}}^\text{SCIPW}\right)-\overline{\Delta}_{k,h}\right\}^{\otimes 2}$, 
with $\overline{\Delta}_{k,h}=n_{k,h}^{-1}\sum_{i\in\mathcal{S}^\mathrm{II}_{k,h}}w_{i}\Delta_{i}\left(\widehat{\pmb{\beta}}^\text{SCIPW}\right)$. The detailed derivation is provided in Supplementary Materials Section 4.1. 

The variance of $\widehat{\mathrm{AUC}}^\text{CIPW}$ can be similarly derived using the total variance formula, while accounting for the randomness due to joint inclusion of case-control pairs, where $\delta_{ij}=\delta_i\cdot\delta_j$ denotes the pairwise inclusion indicator for case $i$ and control $j$. The corresponding joint inclusion probability is $\pi_{ij}=P(\delta_{ij}=1|\mathcal{S}^\mathrm{I})$. As cases and controls are sampled independently under the RMCC design, i.e., $\pi_{ij}=\pi_i \pi_j$, the variance of $\widehat{\mathrm{AUC}}^\text{CIPW}$ can be estimated by 
\begin{equation}
\begin{aligned}
var(\widehat{\mathrm{AUC}}^\text{CIPW}) 
=&\sum_{k=0}^1\frac{N_k}{N_k-1}\sum_{i\in\mathcal{S}^\mathrm{II}_k}\left\{w_i\Delta_i^\text{IPW}-\overline{\Delta}^\text{IPW}_k\right\}^2+\\
&\  var \left( \sum_{i\in\mathcal{S}^\mathrm{II}_1}w_i\Delta_i^\text{CIPW}\middle| \mathcal{S}^\mathrm{I}\right)+var \left( \sum_{i\in\mathcal{S}^\mathrm{II}_0}w_i\Delta_i^\text{CIPW}\middle| \mathcal{S}^\mathrm{I}\right),
\end{aligned}
\end{equation}
where $\Delta_i^\text{IPW}=\partial \widehat{\mathrm{AUC}}^\text{IPW}/\partial w_i$, $\overline{\Delta}^\text{IPW}_k=N_k^{-1}\sum_{i\in\mathcal{S}^\mathrm{II}_k}w_i\Delta_i^\text{IPW}$ for $k=0,1$, and $\Delta_i^\text{CIPW}=\partial\widehat{\mathrm{AUC}}^\text{CIPW}/\partial w_i$. The details of $\Delta_i^\text{IPW}$, $\Delta_i^\text{CIPW}$, along with the complete variance estimation formula, are provided in Supplementary Materials Section 4.2.

\subsection{Replication Variance Estimation for $\widehat{\mathrm{PPV}}^\text{SCIPW}$}

Deriving an influence function–based variance estimator for PPV can be more complex. As an alternative, we use the delete-a-group jackknife (JK) replication method \citep{kim2011,kott1997} to estimate the variance of $\widehat{\mathrm{PPV}}^{\text{SCIPW}}$, accounting for randomness in both the Phase I and Phase II samples.

Let $\mathcal{S}^\mathrm{I}$ be partitioned into $G$ (e.g., $G=10$ in simulation studies in Section 5) disjoint random groups $\{ L_1,...,L_G\}$. For the replicate $g$, we generate the replication weights, denoted by $w_i^{(-g)}$ for $i\in \mathcal{S}^\mathrm{II}$ as follows. The Phase I replication weight is defined as $w_{i}^{\mathrm{I}(-g)}=G(G-1)^{-1}$ for units not in the deleted group and $w_{i}^{\mathrm{I}(-g)}=0$ for units in group $g$. The Phase II replication weight is then defined by $w_{i}^{\mathrm{II}(-g)}=\left\{ \sum_{i\in  \mathcal{S}^\mathrm{II}_{k,h}} w_{i}^{\mathrm{I}(-g)} \right\}^{-1}\left\{\sum_{i\in  \mathcal{S}^\mathrm{I}_{k,h}} w_{i}^{\mathrm{I}(-g)}\right\}$. The final replicate weight for $i\in \mathcal{S}^\mathrm{II}$ is given by $w_i^{(-g)} = w_{i}^{\mathrm{I}(-g)}w_{i}^{\mathrm{II}(-g)}$.

We then re-estimate PPV using $\mathcal{S}^\mathrm{II}$ with the replicate weights $\{w_i^{(-g)},i\in \mathcal{S}^\mathrm{II}\}$ and $\mathcal{S}^\mathrm{I}$ with $\{w_{i}^{\mathrm{I}(-g)},i\in \mathcal{S}^\mathrm{I}\}$ for the calibration step, and denote the resulting replicate estimate by $\widehat{\mathrm{PPV}}^{\text{SCIPW}(-g)}$.
The delete-a-group JK variance estimator for $\widehat{\mathrm{PPV}}^\text{SCIPW}$ is given by
\begin{equation}
var_{JK}\left(\widehat{\mathrm{PPV}}^\text{SCIPW}\right)=\frac{G-1}{G}\sum^G_{g=1}\left( \widehat{\mathrm{PPV}}^{\text{SCIPW}(-g)}-\widehat{\mathrm{PPV}}^\text{SCIPW}\right)^2
\end{equation}
\section{Simulations} 
\subsection{Simulation Settings}

We conducted simulation studies to evaluate the finite-sample performance of the proposed parameter-specific calibration estimators under two-phase designs and to assess the efficiency of the proposed risk-matched case-control (RMCC) sampling design. 
For each simulation replicate, we generated a Phase I sample $\mathcal{S}^{\mathrm{I}}$ of $N=10{,}000$ individuals. The readily available covariate was generated as $X_1\sim N(0,1)$, and the expensive covariate as $X_2\sim\chi^2_8/20$. The binary outcome $Y$ was generated according to $\operatorname{logit}\{P(Y=1\mid X_1,X_2)\}
=\beta_0+\beta_1X_1+\beta_2X_2+\beta_3X_1X_2$. We considered two outcome-model scenarios: a main-effects model with $\beta_3=0$ and an interaction model with $\beta_3=0.2$. Within each scenario, we varied $\beta_0$, $\beta_1$, and $\beta_2$ to assess the effects of three factors on the efficiency of the proposed sampling and estimation procedures: (i) the association between the expensive covariate and the outcome, with $\beta_2\in\{0,0.9\}$; (ii) the discriminatory ability of the final risk model, with AUC values of approximately 0.64, 0.75, and 0.84 corresponding to $\beta_1\in\{0.5,1,1.5\}$; and (iii) outcome prevalence, with event rates of approximately 3\% and 8\%, obtained by varying $\beta_0$.

For each Phase I sample, we selected a Phase II subsample $\mathcal{S}^{\mathrm{II}}$ of size $n=600$ using five sampling designs: the classical case-control (CC) design, the proposed risk-matched case-control (RMCC) design, and three risk-stratified case-control designs based on the same Phase I risk strata as RMCC but using different allocation rules. Specifically, proportional risk-stratified case-control sampling (RPCC) allocates the Phase II sample proportionally across strata, such that ${n_{k,h}}/{N_{k,h}}={n}/{N}$, where $N_{k,h}$ and $n_{k,h}$ denote the Phase I and Phase II sample sizes, respectively, for risk stratum $k$ and outcome stratum $h$. 

Balanced risk-stratified case-control sampling (RBCC) allocates approximately one-third of the cases and one-third of the noncases to each of the three risk strata, i.e., $n_{1,h}\approx n_1/3, n_{0,h}\approx n_0/3$, where $n_h$ denotes the total number of Phase II subjects with outcome $h$.

The Neyman risk-stratified case-control design (RNCC) allocates the Phase II sample across the risk strata in proportion to the stratum-specific contributions to the variance of the estimator of interest. Specifically, for a target parameter $\theta$, the Neyman allocation has the form $n_{k,h}\propto N_{k,h}\sqrt{V_{k,h}(\theta)}$, subject to the constraints on the total numbers of cases and noncases, where $V_{k,h}(\theta)$ denotes the contribution of stratum $(k,h)$ to the variance of the estimator of $\theta$. In the simulations, we set $\theta$ as $\beta_2$ and the stratum-specific variance contributions were calculated using the true data-generating model, so RNCC represents an oracle allocation that provides a benchmark for the efficiency attainable through optimal-stratified allocation for the asymptotic variance of $\beta_2$. 

For each sampling design, we evaluated IPW and the proposed parameter-specific calibration estimators. SCIPW was used for the regression coefficients, whereas CIPW was used for AUC and PPV. Performance was evaluated in terms of empirical bias, variance, relative efficiency, and confidence-interval coverage. All simulations were based on 1,000 independent replicates.

\subsection{Simulation Results}

\begin{figure}[!p]
\centering
    \includegraphics[width=\linewidth]{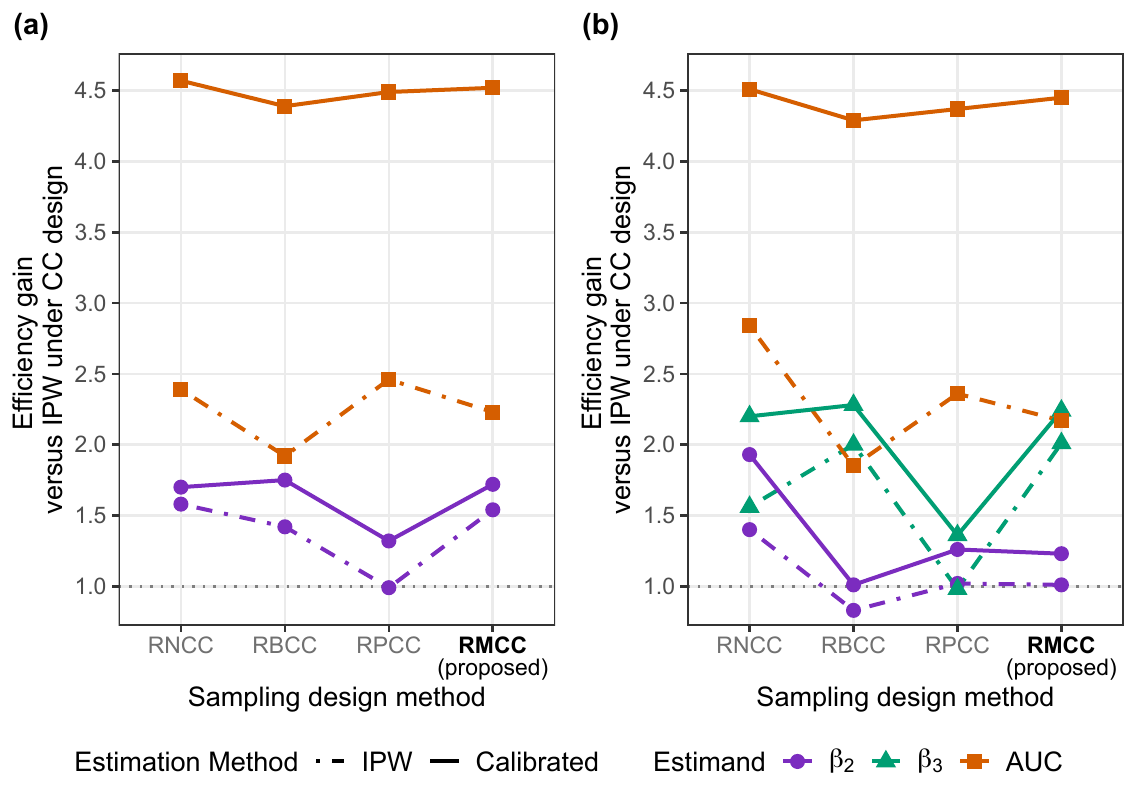}
    \caption{
    Asymptotic relative efficiencies of the Phase II estimators of the regression parameters and AUC under the RNCC, RBCC, RPCC, and RMCC designs, relative to the corresponding IPW estimators under the CC design, for (a) the final risk model without interaction and (b) the final risk model with interaction. Results are shown for an event rate of 8\%, with $\beta_1=1.5$ and $\beta_2=0.9$. The calibrated estimator is parameter-specific: SCIPW is used for the regression parameters $\widehat{\beta}_2$ and $\widehat{\beta}_3$, and CIPW is used for $\widehat{\mathrm{AUC}}$.
    }
\label{fig1}
\end{figure}

We first evaluated the contributions of the proposed sampling designs and parameter-specific calibration estimators to the efficiency of regression coefficient and AUC estimation. The results in Figure \ref{fig1} and Table \ref{tab1} demonstrate that the sampling design and calibration provide complementary efficiency gains for the association parameters $\beta_2, \beta_3$, and AUC. Our proposed RMCC generally achieved the highest or near-highest efficiency across sampling designs, particularly for regression coefficient estimation, while parameter-specific calibration consistently improved efficiency over IPW within each design. The gains from calibration were especially pronounced for AUC estimation, whereas the choice of Phase II sampling design had a greater impact on the efficiency of the regression coefficient estimators. Thus, RMCC combined with SCIPW provided consistently strong performance for regression parameters, while CIPW yielded particularly substantial gains for AUC.

Specifically, Figure \ref{fig1} illustrates these patterns for $\beta_1=1.5$, $\beta_2=0.9$, and an event rate of approximately 8\%. Within each Phase II sampling design, CIPW substantially improves AUC efficiency compared with the corresponding IPW estimator, recovering approximately 85\%--90\% of the efficiency of the corresponding Phase I estimator, compared with approximately 20\%--55\% for IPW. For the regression coefficients, SCIPW also improves efficiency relative to the corresponding IPW estimator under the same sampling design, although the gains are more modest than those observed for AUC.

Table \ref{tab1} demonstrates that these patterns persist across outcome rates, covariate effects, and model specifications. The efficiency advantage of CIPW over IPW for AUC remains under both 3\% and 8\% event rates, while differences among sampling designs are relatively modest. For regression coefficients, sampling design plays a larger role: RMCC generally performs best under an 8\% event rate, whereas RMCC and RBCC are comparable and generally outperform RPCC and RNCC when the event rate is 3\%.

The relative efficiency of the regression coefficient estimators also depends on the target parameter. Without an interaction, the efficiency gain for $\beta_2$ generally increases with $\beta_1$. With an interaction, RMCC provides relatively modest gains for $\beta_2$ but larger gains for $\beta_3$, particularly when $\beta_1$ is large and the outcome is rare. For example, with a 3\% event rate, $\beta_1=1.5$, and $\beta_2=0.9$, the relative efficiency for $\beta_3$ increases from 2.38 with RMCC combined with IPW to 2.74 with RMCC combined with SCIPW.

\begin{table}[!p]
\centering

\caption{
Asymptotic relative efficiencies of the regression coefficient(s) of interest and AUC estimators under the RPCC, RNCC, RBCC, and RMCC designs relative to those under the CC design across scenarios with varied event rates and regression coefficients.
}
\label{tab1}

\begin{threeparttable}

\setlength{\tabcolsep}{1pt}
\renewcommand{\arraystretch}{1.2}
\small

\begin{adjustbox}{max width=\linewidth}
\begin{tabular}{ccrc *{8}{c} c *{8}{c}
}
\toprule


\multicolumn{21}{l}{
\textbf{Without interaction: \(\beta_3=0\)}
}
\\[2pt]
\multicolumn{4}{c}{}& 
\multicolumn{8}{c}{\textbf{RE\textsuperscript{a} versus CC-IPW}
}&&\multicolumn{8}{c}{\textbf{RE versus CC-IPW}}
\\
\cmidrule(lr){5-12}
\cmidrule(lr){14-21}
\textbf{Event}
&&&&
\multicolumn{2}{c}{\textbf{RPCC}}&
\multicolumn{2}{c}{\textbf{RNCC}}&
\multicolumn{2}{c}{\textbf{RMCC}}&
\multicolumn{2}{c}{\textbf{RBCC}}&&
\multicolumn{2}{c}{\textbf{RPCC}}&
\multicolumn{2}{c}{\textbf{RNCC}}&
\multicolumn{2}{c}{\textbf{RMCC}}&
\multicolumn{2}{c}{\textbf{RBCC}}
\\[-3pt]
\cmidrule(lr){5-6}
\cmidrule(lr){7-8}
\cmidrule(lr){9-10}
\cmidrule(lr){11-12}
\cmidrule(lr){14-15}
\cmidrule(lr){16-17}
\cmidrule(lr){18-19}
\cmidrule(lr){20-21}
\textbf{rate}&$\beta_1$&\multicolumn{2}{r}{\textbf{\(\beta_2\)} (var$_{cc}$)\textsuperscript{b}}
&\textbf{IPW}&\textbf{SCIPW}&
\textbf{IPW}&\textbf{SCIPW}&
\textbf{IPW}&\textbf{SCIPW}&
\textbf{IPW}&\textbf{SCIPW}&
\textbf{AUC}(var$_{cc}$)&
\textbf{IPW}&\textbf{CIPW}&
\textbf{IPW}&\textbf{CIPW}&
\textbf{IPW}&\textbf{CIPW}&
\textbf{IPW}&\textbf{CIPW}
\\

\midrule

\multirow{6}{*}{8\%}
& 0.5   & \multicolumn{2}{r}{0.0 (1.92)}
& 1.00 & 1.07
& 1.09 & 1.12
& 1.09 & 1.11
& 1.09 & 1.11
& 0.64 (5.00)
& 3.07 & 4.47
& 3.03 & 4.55
& 3.03 & 4.55
& 2.99 & 4.55
\\

& 1.0   & \multicolumn{2}{r}{0.0 (2.46)}
& 1.00 & 1.20
& 1.34 & 1.39
& 1.33 & 1.39
& 1.31 & 1.39
& 0.75 (3.84)
& 2.87 & 4.77
& 2.70 & 4.81
& 2.67 & 4.81
& 2.49 & 4.78
\\

& 1.5  & \multicolumn{2}{r}{0.0 (3.24)}
& 0.98 & 1.30
& 1.61 & 1.72
& 1.59 & 1.73
& 1.49 & 1.77
& 0.83 (2.64)
& 2.54 & 4.89
& 2.40 & 4.99
& 2.30 & 4.98
& 1.98 & 4.92
\\

& 0.5 & \multicolumn{2}{r}{0.9 (1.86)}
& 1.01 & 1.09
& 1.09 & 1.14
& 1.09 & 1.14
& 1.08 & 1.13
& 0.65 (4.96)
& 2.44 & 3.16
& 2.37 & 3.11
& 2.37 & 3.12
& 2.33 & 3.09
\\

& 1.0 & \multicolumn{2}{r}{0.9 (2.34)}
& 0.99 & 1.20
& 1.31 & 1.38
& 1.29 & 1.38
& 1.26 & 1.38
& 0.76 (3.78)
& 2.63 & 4.06
& 2.52 & 4.10
& 2.45 & 4.07
& 2.29 & 3.99
\\

& 1.5 & \multicolumn{2}{r}{0.9 (3.11)}
& 0.99 & 1.32
& 1.58 & 1.70
& 1.54 & 1.72
& 1.42 & 1.75
& 0.83 (2.61)
& 2.46 & 4.49
& 2.39 & 4.57
& 2.23 & 4.52
& 1.92 & 4.39
\\

\addlinespace
\midrule
\addlinespace

\multirow{6}{*}{3\%}
& 0.5   & \multicolumn{2}{r}{0.0 (1.96)}
& 1.00 & 1.09
& 1.10 & 1.13
& 1.10 & 1.13
& 1.10 & 1.13
& 0.64 (5.06)
& 1.61 & 1.77
& 1.59 & 1.77
& 1.59 & 1.77
& 1.59 & 1.77
\\

& 1.0   & \multicolumn{2}{r}{0.0 (2.67)}
& 0.98 & 1.33
& 1.41 & 1.49
& 1.40 & 1.49
& 1.40 & 1.49
& 0.76 (3.79)
& 1.61 & 1.81
& 1.53 & 1.83
& 1.50 & 1.82
& 1.50 & 1.82
\\

& 1.5   & \multicolumn{2}{r}{0.0 (3.90)}
& 0.98 & 1.67
& 1.90 & 2.06
& 1.84 & 2.12
& 1.84 & 2.13
& 0.84 (2.48)
& 1.57 & 1.85
& 1.45 & 1.86
& 1.35 & 1.86
& 1.34 & 1.86
\\

& 0.5 & \multicolumn{2}{r}{0.9 (1.88)}
& 1.00 & 1.12
& 1.10 & 1.16
& 1.09 & 1.16
& 1.09 & 1.16
& 0.65 (4.99)
& 1.49 & 1.61
& 1.46 & 1.60
& 1.46 & 1.60
& 1.46 & 1.60
\\

& 1.0 & \multicolumn{2}{r}{0.9 (2.53)}
& 0.98 & 1.36
& 1.40 & 1.51
& 1.38 & 1.52
& 1.38 & 1.52
& 0.76 (3.76)
& 1.57 & 1.76
& 1.50 & 1.76
& 1.45 & 1.75
& 1.45 & 1.74
\\

& 1.5 & \multicolumn{2}{r}{0.9 (3.71)}
& 0.99 & 1.70
& 1.86 & 2.06
& 1.78 & 2.13
& 1.77 & 2.13
& 0.85 (2.45)
& 1.55 & 1.82
& 1.46 & 1.83
& 1.33 & 1.81
& 1.32 & 1.81
\\


\midrule
\multicolumn{21}{l}{
\textbf{With interaction: \(\beta_3=0.2\)}
}
\\[2pt]
\multicolumn{3}{c}{}& 
\multicolumn{8}{c}{\textbf{RE versus CC-IPW}
}&&\multicolumn{8}{c}{\textbf{RE versus CC-IPW}}
\\
\cmidrule(lr){5-12}
\cmidrule(lr){14-21}
\textbf{Event}
&&&&
\multicolumn{2}{c}{\textbf{RPCC}}&
\multicolumn{2}{c}{\textbf{RNCC}}&
\multicolumn{2}{c}{\textbf{RMCC}}&
\multicolumn{2}{c}{\textbf{RBCC}}&&
\multicolumn{2}{c}{\textbf{RPCC}}&
\multicolumn{2}{c}{\textbf{RNCC}}&
\multicolumn{2}{c}{\textbf{RMCC}}&
\multicolumn{2}{c}{\textbf{RBCC}}
\\[-3pt]
\cmidrule(lr){5-6}
\cmidrule(lr){7-8}
\cmidrule(lr){9-10}
\cmidrule(lr){11-12}
\cmidrule(lr){14-15}
\cmidrule(lr){16-17}
\cmidrule(lr){18-19}
\cmidrule(lr){20-21}
\textbf{rate}&$\beta_1$&\multicolumn{2}{r}{\textbf{\(\beta_2\)} (var$_{cc}$)}
&\textbf{IPW}&\textbf{SCIPW}&
\textbf{IPW}&\textbf{SCIPW}&
\textbf{IPW}&\textbf{SCIPW}&
\textbf{IPW}&\textbf{SCIPW}&
\textbf{AUC}(var$_{cc}$)&
\textbf{IPW}&\textbf{CIPW}&
\textbf{IPW}&\textbf{CIPW}&
\textbf{IPW}&\textbf{CIPW}&
\textbf{IPW}&\textbf{CIPW}
\\

\midrule

\multirow{6}{*}{8\%}
& 0.5   & \multicolumn{2}{r}{0.0 (2.08)}
& 1.01 & 1.04
& 1.11 & 1.17
& 0.96 & 0.99
& 0.92 & 0.96
& 0.66 (4.86)
& 3.02 & 4.37
& 3.12 & 4.34
& 2.96 & 4.44
& 2.89 & 4.38
\\

& 1.0   & \multicolumn{2}{r}{0.0 (3.04)}
& 1.00 & 1.10
& 1.26 & 1.51
& 0.95 & 1.04
& 0.86 & 0.94
& 0.77 (3.62)
& 2.76 & 4.62
& 3.09 & 4.65
& 2.57 & 4.66
& 2.38 & 4.64
\\

& 1.5   & \multicolumn{2}{r}{0.0 (5.14)}
& 1.01 & 1.24
& 1.42 & 1.93
& 1.05 & 1.23
& 0.86 & 1.02
& 0.84 (2.44)
& 2.45 & 4.78
& 2.91 & 4.89
& 2.22 & 4.85
& 1.89 & 4.77
\\

& 0.5 & \multicolumn{2}{r}{0.9 (1.98)}
& 0.99 & 1.05
& 1.10 & 1.18
& 0.93 & 1.00
& 0.90 & 0.96
& 0.67 (4.78)
& 2.44 & 3.23
& 2.51 & 3.22
& 2.39 & 3.26
& 2.34 & 3.21
\\

& 1.0 & \multicolumn{2}{r}{0.9 (2.85)}
& 1.01 & 1.13
& 1.24 & 1.51
& 0.92 & 1.04
& 0.82 & 0.94
& 0.77 (3.55)
& 2.54 & 3.99
& 2.84 & 4.07
& 2.37 & 4.01
& 2.19 & 3.92
\\

& 1.5 & \multicolumn{2}{r}{0.9 (4.71)}
& 1.02 & 1.26
& 1.40 & 1.93
& 1.01 & 1.23
& 0.83 & 1.01
& 0.85 (2.41)
& 2.36 & 4.37
& 2.84 & 4.51
& 2.17 & 4.45
& 1.85 & 4.29
\\

\addlinespace
\midrule
\addlinespace

\multirow{6}{*}{3\%}
& 0.5   & \multicolumn{2}{r}{0.0 (2.20)}
& 1.00 & 1.04
& 1.03 & 1.08
& 0.92 & 0.96
& 0.92 & 0.95
& 0.66 (4.88)
& 1.61 & 1.76
& 1.61 & 1.76
& 1.58 & 1.77
& 1.58 & 1.77
\\

& 1.0   & \multicolumn{2}{r}{0.0 (3.65)}
& 1.02 & 1.19
& 1.10 & 1.24
& 0.93 & 1.02
& 0.92 & 1.02
& 0.77 (3.56)
& 1.60 & 1.81
& 1.61 & 1.82
& 1.48 & 1.82
& 1.47 & 1.82
\\

& 1.5   & \multicolumn{2}{r}{0.0 (7.06)}
& 1.01 & 1.43
& 1.30 & 1.53
& 1.04 & 1.24
& 1.03 & 1.24
& 0.86 (2.27)
& 1.56 & 1.86
& 1.58 & 1.88
& 1.33 & 1.87
& 1.32 & 1.87
\\

& 0.5 & \multicolumn{2}{r}{0.9 (2.09)}
& 1.01 & 1.07
& 1.05 & 1.12
& 0.91 & 0.98
& 0.91 & 0.98
& 0.67 (4.80)
& 1.49 & 1.62
& 1.48 & 1.61
& 1.45 & 1.61
& 1.45 & 1.61
\\

& 1.0 & \multicolumn{2}{r}{0.9 (3.38)}
& 1.02 & 1.22
& 1.11 & 1.29
& 0.88 & 1.01
& 0.87 & 1.01
& 0.78 (3.51)
& 1.55 & 1.74
& 1.57 & 1.76
& 1.42 & 1.74
& 1.42 & 1.73
\\

& 1.5 & \multicolumn{2}{r}{0.9 (6.52)}
& 1.00 & 1.46
& 1.30 & 1.58
& 0.98 & 1.23
& 0.98 & 1.22
& 0.86 (2.24)
& 1.54 & 1.81
& 1.56 & 1.83
& 1.31 & 1.81
& 1.30 & 1.81
\\




\midrule

\multicolumn{5}{c}{}
&
\multicolumn{8}{c}{
\textbf{RE versus CC-IPW}
}
&
\multicolumn{8}{c}{}
\\[-1pt]
\cmidrule(lr){6-13}
\textbf{Event}&&&&&
\multicolumn{2}{c}{\textbf{RPCC}}&
\multicolumn{2}{c}{\textbf{RNCC}}&
\multicolumn{2}{c}{\textbf{RMCC}}&
\multicolumn{2}{c}{\textbf{RBCC}}&
\multicolumn{8}{c}{}
\\[-3pt]
\cmidrule(lr){6-7}
\cmidrule(lr){8-9}
\cmidrule(lr){10-11}
\cmidrule(lr){12-13}

\textbf{rate}&
\textbf{\(\beta_1\)}
&&
\textbf{\(\beta_2\)} & var$_{cc}(\widehat{\beta}_3)$&
\textbf{IPW}&\textbf{SCIPW}&
\textbf{IPW}&\textbf{SCIPW}&
\textbf{IPW}&\textbf{SCIPW}&
\textbf{IPW}&\textbf{SCIPW}&
\multicolumn{8}{c}{}
\\
\midrule

\multirow{6}{*}{8\%}
& 0.5  && 0.0 &2.39
& 1.02 & 1.14
& 0.87 & 1.17
& 1.22 & 1.28
& 1.23 & 1.28
& \multicolumn{8}{c}{}
\\

& 1.0  && 0.0 &3.65
& 0.99 & 1.25
& 1.00 & 1.62
& 1.61 & 1.72
& 1.65 & 1.75
& \multicolumn{8}{c}{}
\\

& 1.5  && 0.0 & 5.89
& 0.98 & 1.35
& 1.65 & 2.29
& 2.02 & 2.22
& 2.03 & 2.27
& \multicolumn{8}{c}{}
\\

& 0.5 && 0.9 & 2.28
& 1.00 & 1.13
& 0.88 & 1.16
& 1.18 & 1.26
& 1.19 & 1.26
& \multicolumn{8}{c}{}
\\

& 1.0 && 0.9 & 3.49
& 0.99 & 1.25
& 0.92 & 1.55
& 1.56 & 1.70
& 1.58 & 1.71
& \multicolumn{8}{c}{}
\\

& 1.5 && 0.9 & 5.65
& 0.98 & 1.36
& 1.56 & 2.20
& 2.01 & 2.24
& 2.00 & 2.28
& \multicolumn{8}{c}{}
\\

\addlinespace
\midrule
\addlinespace

\multirow{6}{*}{3\%}
& 0.5   && 0.0 & 2.50
& 0.99 & 1.16
& 0.88 & 1.14
& 1.26 & 1.32
& 1.26 & 1.32
& \multicolumn{8}{c}{}
\\

& 1.0   && 0.0 &4.00
& 1.00 & 1.46
& 1.32 & 1.74
& 1.84 & 2.00
& 1.85 & 2.01
& \multicolumn{8}{c}{}
\\

& 1.5   && 0.0 & 6.41
& 0.98 & 1.69
& 2.20 & 2.54
& 2.48 & 2.77
& 2.48 & 2.76
& \multicolumn{8}{c}{}
\\

& 0.5 && 0.9 & 2.37
& 1.01 & 1.20
& 0.88 & 1.17
& 1.25 & 1.34
& 1.24 & 1.34
& \multicolumn{8}{c}{}
\\

& 1.0 && 0.9 & 3.62
& 0.99 & 1.42
& 1.21 & 1.65
& 1.69 & 1.90
& 1.69 & 1.90
& \multicolumn{8}{c}{}
\\

& 1.5 && 0.9 &6.12
& 0.99 & 1.69
& 2.08 & 2.48
& 2.38 & 2.74
& 2.38 & 2.74
& \multicolumn{8}{c}{}
\\

\bottomrule

\end{tabular}
\end{adjustbox}

\vspace{2mm}

\begin{minipage}{0.94\linewidth}
\footnotesize

\textsuperscript{a}
Relative efficiency: the ratio of the asymptotic variance
of the CC-IPW estimator to the asymptotic variance of the corresponding
Phase II estimator. Values greater than one indicate improved efficiency
relative to CC-IPW.

\textsuperscript{b}
var$_{cc}$ is the simulation mean of the asymptotic
variance of the corresponding CC-IPW estimator ($\times 10$ for $\widehat{\beta}_2$ and $\widehat{\beta}_3$; $\times 10^{4}$ for $\widehat{\text{AUC}}$ ). 
\end{minipage}
\end{threeparttable}
\end{table}
The Phase II sample allocations help explain these efficiency patterns (Tables S1 and S2 in Supplementary Materials).
At the 3\% event rate, RMCC produced nearly identical allocations to RBCC,
whereas at the 8\% event rate, its allocations were generally closer to
those of RNCC in models without an interaction. When an interaction was
present, RNCC, which targets minimization of the variance of
$\widehat{\beta}_2$ alone, produced substantially more uneven allocations
across risk strata. These allocations favored estimation of $\beta_2$
but generally yielded smaller efficiency gains for $\beta_3$, particularly
under IPW. In contrast, RMCC allocated the Phase II sample more evenly
across risk strata, generally improving efficiency for $\beta_3$ while
maintaining efficiency for $\beta_2$ close to or above that of the CC-IPW
benchmark when combined with SCIPW. Thus, RMCC offered a compromise
between efficient estimation of the main effect and the interaction. Efficiency gains for estimating the effect of the inexpensive covariate,
$\beta_1$, were generally more substantial than those for the parameters
involving the expensive covariate, $\beta_2$ and $\beta_3$, although the
overall patterns across sampling designs and estimators were similar (Tables S3 and S4 in Supplementary Materials).

Table \ref{tab2} reports the proposed analytical variance estimates for the SCIPW estimators of parameters of interests under the proposed RMCC design. The ratios of the analytical to empirical variance estimates were generally close to one for $\beta_2$, $\beta_3$, and AUC, and the corresponding 95\% coverage probabilities were close to the nominal level. These results indicate a good performance of the proposed variance estimators and the validity of asymptotic normal inference under the RMCC design.

\begin{table}[!p]
\centering
\caption{
Asymptotic variances, variance ratios, and 95\% coverage probabilities under RMCC design, using SCIPW for the regression coefficients and CIPW for AUC, in models with and without an interaction.
}
\label{tab2}

\begin{threeparttable}

\small
\renewcommand{\arraystretch}{1.20}

\begin{tabular*}{\linewidth}{
    @{\extracolsep{\fill}}
    c c c c c c c
}
\toprule

\multicolumn{7}{l}{
\textbf{Without an interaction: \(\beta_3=0\)}
}
\\[2pt]

\textbf{Event rate}
&
\textbf{\(\beta_1\)}
&
\makecell{\textbf{\(\beta_2\)}\textbf{(Phase I\textsuperscript{a})}}
&
\makecell{\textbf{\(\operatorname{var}(\widehat{\beta}_2)\)}\textsuperscript{b}\textbf{(VR\textsuperscript{c})}}
&
\textbf{Coverage}\textsuperscript{d}
&
\makecell{\textbf{\(\operatorname{var}(\widehat{\mathrm{AUC}})\)}\textbf{(VR)}}
&
\textbf{Coverage}
\\
\midrule

\multirow{6}{*}{8\%}
& 0.5 & 0.0 (3.33) & 1.73 (1.01) & 0.95 & 1.10 (0.97) & 0.94 \\
& 0.5 & 0.9 (2.91) & 1.64 (1.00) & 0.96 & 1.59 (1.03) & 0.95 \\
& 1.0 & 0.0 (3.68) & 1.77 (0.99) & 0.96 & 0.80 (1.01) & 0.94 \\
& 1.0 & 0.9 (3.29) & 1.70 (1.03) & 0.96 & 0.93 (1.02) & 0.95 \\
& 1.5 & 0.0 (4.11) & 1.87 (0.92) & 0.95 & 0.53 (1.02) & 0.95 \\
& 1.5 & 0.9 (3.73) & 1.80 (1.02) & 0.95 & 0.58 (1.07) & 0.95 \\

\addlinespace[2pt]
\midrule
\addlinespace[2pt]

\multirow{6}{*}{3\%}
& 0.5 & 0.0 (9.37)  & 1.74 (0.93) & 0.94 & 2.86 (0.91) & 0.94 \\
& 0.5 & 0.9 (7.91)  & 1.63 (1.05) & 0.96 & 3.12 (0.94) & 0.95 \\
& 1.0 & 0.0 (9.60)  & 1.79 (0.96) & 0.95 & 2.08 (0.94) & 0.94 \\
& 1.0 & 0.9 (8.25)  & 1.67 (0.92) & 0.94 & 2.15 (0.98) & 0.95 \\
& 1.5 & 0.0 (10.30) & 1.84 (0.95) & 0.95 & 1.33 (1.12) & 0.96 \\
& 1.5 & 0.9 (8.97)  & 1.74 (0.98) & 0.95 & 1.35 (1.06) & 0.95 \\

\end{tabular*}

\vspace{2mm}

\noindent\makebox[\linewidth]{\rule{\linewidth}{0.4pt}}

\vspace{2mm}

\begin{adjustbox}{max width=\linewidth}

\setlength{\tabcolsep}{3.5pt}
\renewcommand{\arraystretch}{1.18}

\begin{tabular}{
    c c c c c c c c c
}

\multicolumn{9}{l}{
    \textbf{With an interaction: \(\beta_3=0.2\)}
}
\\[2pt]

\textbf{Event rate}
&
\textbf{\(\beta_1\)}
&
\makecell{\textbf{\(\beta_2\)}\textbf{(Phase I)}}
&
\makecell{\textbf{\(\operatorname{var}(\widehat{\beta}_2)\)}\textbf{(VR)}}
&
\textbf{Coverage}
&
\makecell{\textbf{\(\operatorname{var}(\widehat{\beta}_3)\)}\textbf{(VR)}}
&
\textbf{Coverage}
&
\makecell{\textbf{\(\operatorname{var}(\widehat{\mathrm{AUC}})\)}\textbf{(VR)}}
&
\textbf{Coverage}
\\
\midrule

\multirow{6}{*}{8\%}
& 0.5 & 0.0 (4.24)  & 2.09 (0.98) & 0.95 & 1.87 (0.97) & 0.96 & 1.09 (0.99) & 0.95 \\
& 0.5 & 0.9 (3.69)  & 1.99 (1.04) & 0.96 & 1.81 (0.97) & 0.94 & 1.47 (0.97) & 0.94 \\
& 1.0 & 0.0 (6.64)  & 2.91 (0.97) & 0.95 & 2.12 (0.97) & 0.95 & 0.78 (1.01) & 0.94 \\
& 1.0 & 0.9 (5.75)  & 2.73 (0.98) & 0.95 & 2.05 (1.05) & 0.96 & 0.89 (1.01) & 0.95 \\
& 1.5 & 0.0 (10.51) & 4.16 (0.95) & 0.95 & 2.65 (1.01) & 0.96 & 0.50 (1.06) & 0.95 \\
& 1.5 & 0.9 (9.04)  & 3.82 (0.93) & 0.94 & 2.52 (0.99) & 0.95 & 0.54 (1.02) & 0.95 \\

\addlinespace[2pt]
\midrule
\addlinespace[2pt]

\multirow{6}{*}{3\%}
& 0.5 & 0.0 (12.29) & 2.30 (1.02) & 0.96 & 1.90 (0.97) & 0.95 & 2.75 (0.94) & 0.94 \\
& 0.5 & 0.9 (10.42) & 2.14 (1.07) & 0.96 & 1.77 (0.97) & 0.95 & 2.98 (0.96) & 0.94 \\
& 1.0 & 0.0 (19.26) & 3.57 (0.96) & 0.94 & 2.00 (0.98) & 0.95 & 1.96 (0.94) & 0.94 \\
& 1.0 & 0.9 (16.40) & 3.34 (1.00) & 0.95 & 1.91 (0.97) & 0.94 & 2.02 (0.94) & 0.95 \\
& 1.5 & 0.0 (31.63) & 5.69 (0.97) & 0.95 & 2.32 (0.97) & 0.95 & 1.21 (1.13) & 0.95 \\
& 1.5 & 0.9 (27.24) & 5.30 (0.98) & 0.95 & 2.23 (1.02) & 0.95 & 1.24 (1.11) & 0.96 \\

\bottomrule
\end{tabular}
\end{adjustbox}

\vspace{2mm}

\begin{minipage}{0.94\linewidth}
\footnotesize

\textsuperscript{a}
Phase I reports the mean estimated asymptotic variance of the corresponding full cohort estimator ($\times 10^2$).

\textsuperscript{b}
var is the simulation mean of the asymptotic
variance of the corresponding RMCC SCIPW estimator ($\times 10$ for $\widehat{\beta}_2$ and $\widehat{\beta}_3$; $\times 10^{4}$ for $\widehat{\text{AUC}}$ ).

\textsuperscript{c}
VR is the ratio of the mean estimated asymptotic variance to the empirical variance. Values close to one indicate agreement between two variance measures.

\textsuperscript{d}
The 95\% coverage probability is the proportion of simulation replicates in which the asymptotic normal 95\% confidence interval contains the true parameter value.

\end{minipage}

\end{threeparttable}
\end{table}

\begin{figure}[!p]
    \centering
    \includegraphics[width=1.05\linewidth]{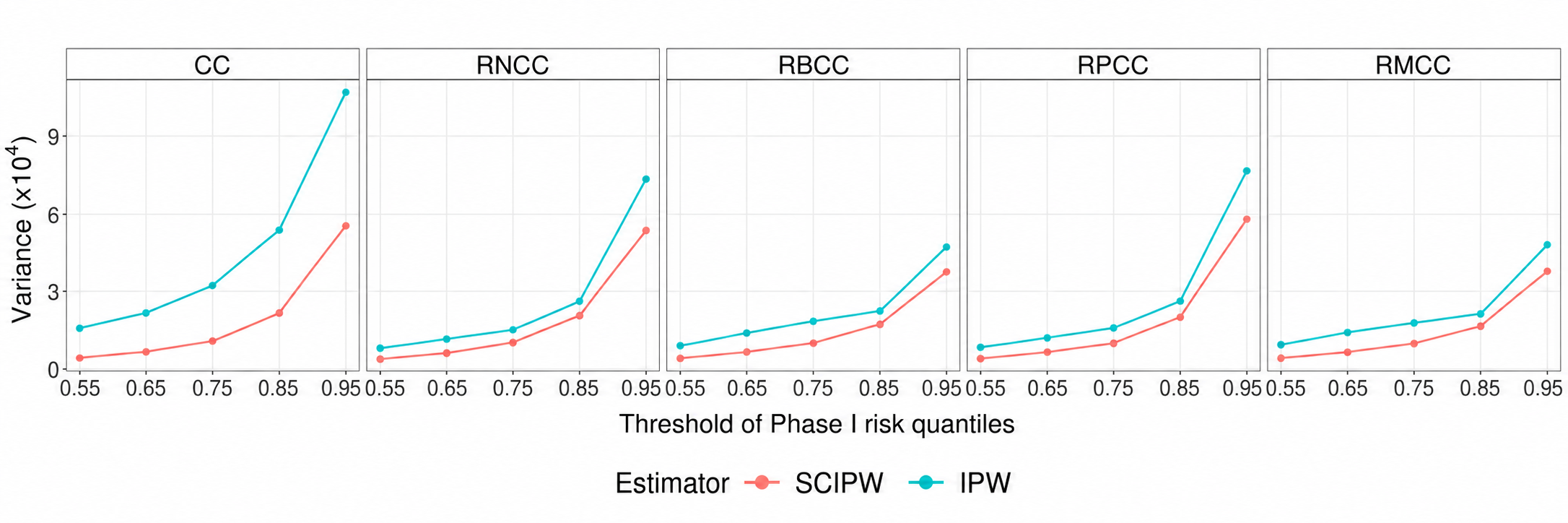}
    \caption{Empirical variances of the PPV estimators across Phase I risk-quantile thresholds under the RNCC, RBCC, RPCC, and RMCC designs for the setting with event rate of 8\% and $(\beta_1, \beta_2, \beta_3)^{\prime}=(1.5, 0.9, 0.2)^{\prime}$. PPV is evaluated at thresholds corresponding to the 55th, 65th, 75th, 85th, and 95th percentile of the Phase I predicted risk distribution. 
   Within each design, IPW and SCIPW estimators are compared. RPCC, RNCC, RBCC denote proportional, Neyman, and balanced risk-stratified case-control designs, respectively, while RMCC denotes the proposed risk-matched case-control design.} \label{fig2}
\end{figure}

Finally, Figure \ref{fig2} shows how the efficiency of PPV estimation varies with the risk threshold $c$, defined by Phase I risk percentiles ranging from the 55th to the 95th percentile in 10-percentage-point increments. 
Across all sampling designs, the analytical variance increased as the threshold moved toward the upper tail of the risk distribution, reflecting the smaller number of individuals classified as high risk at more extreme thresholds. SCIPW consistently yielded lower variance than IPW, with the reduction particularly pronounced at the 85th and 95th percentiles. Among the sampling designs, RMCC generally achieved the highest efficiency at these extreme thresholds, with performance comparable to RBCC. Therefore, the proposed calibration and RMCC procedures can improve efficiency for PPV estimation, especially when inference targets small, high-risk subgroups.

\section{Application to Oral Precancer Progression Risk Prection Using Data from a Taiwan Cohort Study }

We illustrate the proposed methodology using data from a cohort study of patients with oral precancer in Taiwan, led by the National Cancer Institute. Between 2013 and 2019, the study enrolled 1,998 patients with clinically diagnosed oral precancerous lesions and followed them biannually for up to five years to characterize the natural history of precancer, including regression, persistence, incidence, and progression. Participants were also passively followed through periodic linkage with Taiwan\rq s national cancer registries. Questionnaire data and biospecimens were collected at multiple follow-up visits and stored for future analyses. Additional details on the study design and data collection are provided in \citet{wangcp2025}.

For this analysis, we define the binary outcome $Y$ as cancer progression (CP) during the follow-up period and develop a risk prediction model using baseline covariates. Approximately 3.2\% of patients developed oral cancer during the follow up. The Phase I covariates $\pmb{X}_1$ include age in 10 years (Age), sex (Sex), education (Educ) with 7 levels, ranging from illiterate to college/university above, and a three-level categorical summary score for three health-related behaviors including smoking cigarettes, drinking alcohol, and chewing betel quid. Each individual is assigned a score for each behavior (0, 1, and 2 for never, former, and current users, respectively). The three scores are summed, resulting in a total score ranging from 0 to 6, and then categorized as \lq\lq no risky behaviors\rq\rq\ (NoRB) with score$=$0, \lq\lq some risky behaviors\rq\rq\ (SomeRB) with score between 1 and 4, and \lq\lq severe risky behaviors\rq\rq\ (SevRB) with score$>$4. To account for the duration of any risky behaviors  and avoid collinearity issue due to its strong correlation with age, we include the effect of current age (Age) for individuals with no unhealthy behaviors and age at initiation of smoking, drinking, or betel quid chewing (AgeStart) for individuals who have any unhealthy behaviors (AnyRB, with score$>$0), defined as current age minus the duration of the earliest risky behavior.

We consider the baseline lesion pathological score as the expensive Phase II covariate \(X_2\). This score summarizes the histological severity of the oral lesion and ranges from 0 to 3: 0 indicates no dysplasia, including lesions with oral submucosal fibrosis or lichen planus but no dysplasia; scores of 1, 2, and 3 indicate mild, moderate, and severe dysplasia, respectively. It captures tissue-level information that may contribute to predicting CP beyond the demographic and behavioral characteristics available in Phase I. Obtaining the score requires an invasive biopsy, tissue processing, and expert histopathologic evaluation, making its collection costly and time consuming. The full model of interest is then 
\begin{align*}
\text{logit}\left\{\Pr(CP=1|\pmb{X})\right\}=
&\beta_{10}+\beta_{11}\text{Sex}+\beta_{12}\text{Educ}+\beta_{13}\text{SomeRB}+\beta_{14}\text{SevRB}+\\
&\beta_{15}(\text{Age}\times\text{NoRB})+\beta_{16}(\text{AgeStart}\times\text{AnyRB})+\beta_2\text{PathScore}. 
\label{CPRiskModel}
\end{align*}
Although the pathological score was measured for every individual in this cohort, we emulate a two-phase design by restricting its availability during sampling and analysis. Specifically, Phase II selection is conducted without using the pathological score, and subsequent analyses use its values only for individuals selected into the Phase II subsample; values for all other individuals are masked and treated as missing. The model fitted to the complete Phase I cohort, including the pathological score, provides a full-cohort benchmark for evaluating the proposed sampling design and estimation methods.

In the sampling stage, we implement the proposed RMCC to select a Phase II sample with two sample sizes ($n$=280 and $n$=500). We  partition the Phase I sample into $H=3$ risk-matching strata, with the cut points $(c_1, c_2)$ defined as the empirical tertiles of the estimated risks among Phase I cases, estimated from the Phase I primary risk model $\text{logit}\{\Pr(CP=1 \mid \pmb{X}_1)\} = \alpha_0 + \pmb{\alpha}_1^\top\pmb{X}_1$. We then compare association parameter estimates and predictive performance across three analysis settings: (1) RMCC with IPW, (2) RMCC with the proposed weight calibration, and (3) CC with IPW. For reference, we also report results from a full data (FULL) analysis, in which all individuals are included in Phase II (i.e. $n=1,998$). 

\begin{figure}[!p]
\centering
    \includegraphics[width=\linewidth]{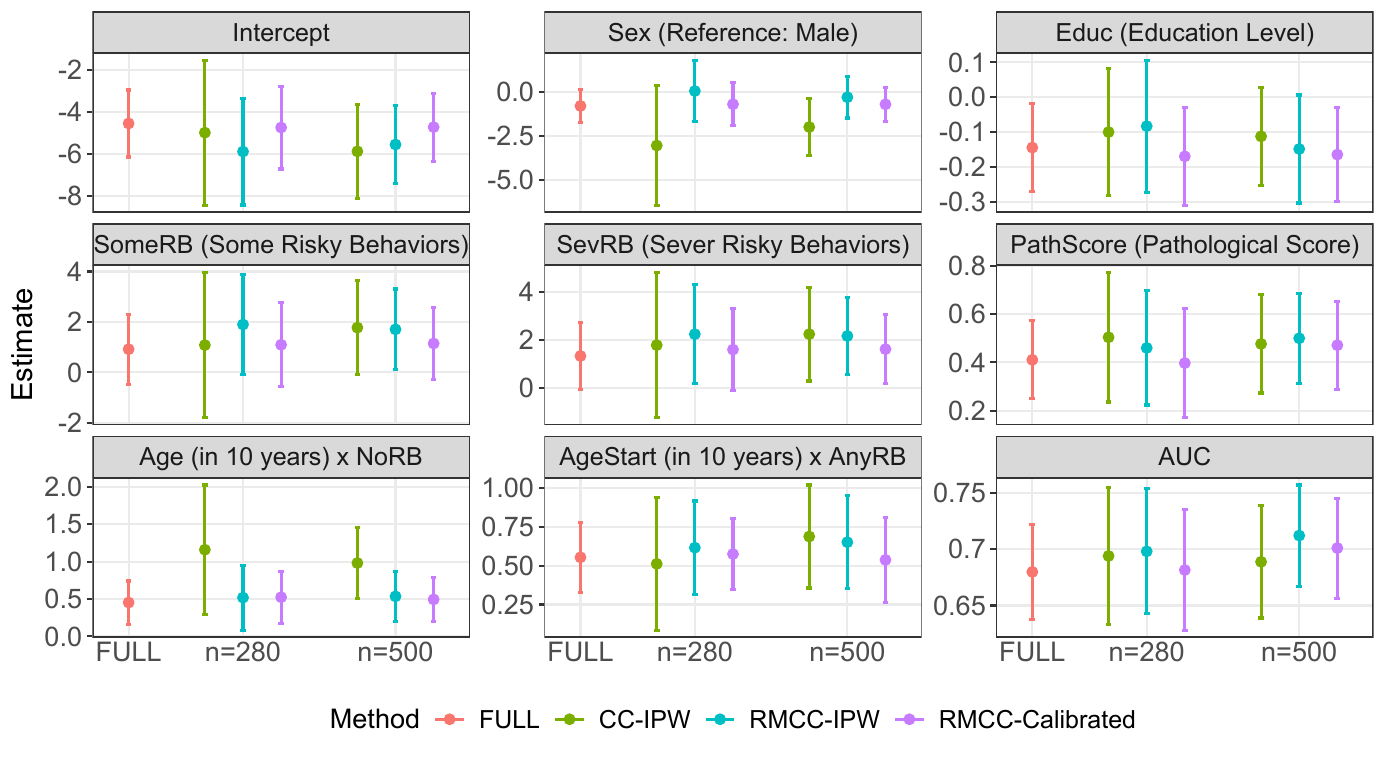}
    \caption{Point estimates and 95\% confidence intervals for regression coefficients and AUC for the analysis model under RMCC and CC Phase II sampling designs compared with FULL Phase I cohort.}
\label{fig3}
\end{figure}

The results are summarized in Figure \ref{fig3} and Tables S5 and S6 in the Supplementary Materials. Across Phase II sample sizes, the point estimates under RMCC are generally comparable to those from the FULL analysis, with the primary differences arising in estimation precision. 
Consistent with the simulation findings, our proposed method, combining the RMCC design with calibrated weights, provides substantial efficiency gains over IPW estimators under the conventional CC design for estimating both association parameters and AUC.

For \(\beta_2\), SCIPW under the RMCC design achieves approximately 51.3\% of the efficiency of the FULL analysis when \(n=280\), while being 40\% more efficient than IPW under the CC design and 10\% more efficient than IPW under the RMCC design. For \(\pmb{\beta}_1=(\beta_{10}, \beta_{11}, \beta_{12}, \beta_{13}, \beta_{14}, \beta_{15}, \beta_{16})^{\top}\), SCIPW under the RMCC design provides substantial efficiency gains relative to IPW under the CC design, with efficiency ratios of up to 7.7 for sex and 6.1 for current age among participants with no risky behaviors. For AUC, the combination of the RMCC design and weight calibration estimation is up to 1.35 times as efficient as CC with IPW. As expected, increasing the Phase II sample size from \(n=280\) to \(n=500\) improves the efficiency of all Phase II designs and estimation methods and results in narrower confidence intervals for all parameter estimates. The relative efficiency patterns remain broadly similar across designs and estimation methods at the two Phase II sample sizes.

From this data application, we observe that, had our proposed design and analysis framework been adopted to assess pathological scores in only a subset of patients at recruitment, pathological assessment could potentially have been reduced while maintaining comparable inferential precision, with corresponding potential savings in study costs. More broadly, the Taiwan cohort provides rich biospecimen resources for identifying clinicopathologic factors and validating biomarkers for predicting cancer progression, making it particularly well suited for illustrating the potential value of the proposed framework in biomarker investigations.

\section{Discussion}
We develop a general framework for improving the efficiency of estimating multiple target parameters in two-phase studies of binary outcomes by incorporating Phase I information into both sampling design and inference. At the sampling stage, we propose an RMCC design that defines sampling strata using the outcome and predicted risks from a Phase I risk model, allocates case sample sizes across risk strata according to the primary risk residuals, and selects controls so that the proportions across risk strata match those of the selected cases. The resulting Phase II subsample approximately preserves the weighted distribution of primary risk among Phase I cases while oversampling individuals with large primary risk residuals for both cases and controls. At the inference stage, we propose target-specific weight calibration methods for estimating association parameters, PPV, and AUC. These methods adjust the original inverse probability sampling weights using information from the corresponding parameters estimated under the Phase I risk model.

Across the simulation studies and data application, the proposed RMCC design combined with the SCIPW approach generally yielded the most efficient estimators of the association parameters while maintaining design consistency. The SCIPW estimator also substantially improved the efficiency of PPV estimation when the weights were calibrated to the association parameters, but provided limited gains for AUC estimation. In contrast, calibrating the weights using the AUC estimator from the Phase I risk model led to substantial efficiency gains for AUC estimation under the Phase II risk model. These findings suggest that the efficiency gains from weight calibration depend on the alignment between the calibration variables and the target parameters, highlighting the importance of tailoring the calibration procedure to the parameter or prediction measure of interest.

Some extensions of the proposed methods merit further investigation. First, our simulation studies consider settings in which no informative surrogate of $X_2$ is available in Phase I. As expected, the proposed methods yield more substantial efficiency gains for estimating $\pmb{\beta}_1$, the effects of the Phase I covariates $\pmb{X}_1$, than for estimating $\beta_2$, the effect of the expensive Phase II covariate $X_2$. This finding is consistent with those reported in the literature \citep{chen2022, wang2025}. When an informative surrogate of $X_2$, denoted by $X_2^*$, is available in Phase I, it can be incorporated into the Phase I risk model as a proxy for $X_2$. A more informative Phase I risk score can improve efficiency at two stages. At the sampling stage, it can better distinguish individuals according to their underlying disease risk and thereby facilitate more informative allocation of the Phase II subsampling probabilities. At the inference stage, the surrogate provides additional Phase I information that can be incorporated into the estimation procedure through calibration, potentially reducing residual variation and improving the precision of $\hat{\beta}_2$.

Second, the proposed weight scaling approach relies on a correct specification of the final logistic model. Although weight scaling can improve the efficiency of the weight calibration estimator for regression parameters by borrowing strength from the specified risk model, it may also introduce bias when the final risk model is misspecified \citep{scott1986}. Future work could investigate alternative weight adjustment approaches, such as weight smoothing \citep{beaumont2008,kim2013}, that are less sensitive to misspecification of the outcome model, as well as extend the scaling approach to other generalized linear models.

Third, the current framework focus on binary outcomes and logistic regression risk models. Extending the RMCC design and associated design-based inference to time-to-event outcomes are of great interests. Future research could investigate RMCC designs based on Cox regression risk models while accounting for censoring and other complexities of time-to-vent outcomes.

\section*{Acknowledgement}
The authors gratefully acknowledge Drs. Anil Chaturvedi and Kelly Yu for sharing the data from the oral precancer natural history study in Taiwan.
The authors also thank Drs. Barry Graubard, Li Cheung, and Anil Chaturvedi for their valuable comments and insightful suggestions on the methodological development and data application of this project.

\section*{Data Availability}
The data used in this study were provided by Dr. Anil Chaturvedi of the Division of Cancer Epidemiology and Genetics, National Cancer Institute. Requests for access to the data should be directed to Dr. Chaturvedi (chaturva@mail.nih.gov). 
Simulation codes are available at https://github.com/jzhang419/two-phase-binary-outcome-simulation.

\bibliographystyle{biom} 
\bibliography{references.bib}

\end{document}